\documentclass[aip,apl,amsmath,amssymb,reprint,floatfix]{revtex4-2}

\usepackage{graphicx}
\usepackage{bm}
\usepackage{xcolor}

\graphicspath{{Figures/}{../analysis/}}

\begin{document}

\title{Coherent antiferromagnetic resonance in MnO driven by impulsive terahertz excitation}

\author{Yinchuan Lv}
\affiliation{Stanford Institute for Materials and Energy Sciences, SLAC National Accelerator Laboratory, 2575 Sand Hill Road, Menlo Park, California 94025, USA}
\email{lvyc@stanford.edu}
\author{Martin J. Cross}
\affiliation{SLAC National Accelerator Laboratory, 2575 Sand Hill Road, Menlo Park, California 94025, USA}
\author{Hari Paudyal}
\affiliation{Department of Physics and Astronomy, University of Iowa, Iowa City, Iowa 52242, USA}
\author{
Christopher T. Parzyck}
\affiliation{Stanford Institute for Materials and Energy Sciences, SLAC National Accelerator Laboratory, 2575 Sand Hill Road, Menlo Park, California 94025, USA}
\author{A. H. M. Reid}
\affiliation{SLAC National Accelerator Laboratory, 2575 Sand Hill Road, Menlo Park, California 94025, USA}
\author{Durga Paudyal}
\affiliation{Department of Physics and Astronomy, University of Iowa, Iowa City, Iowa 52242, USA}
\author{Matthias C. Hoffmann}
\affiliation{SLAC National Accelerator Laboratory, 2575 Sand Hill Road, Menlo Park, California 94025, USA}

\date{\today}

\begin{abstract}
Antiferromagnets (AFMs) offer a promising platform for ultrafast information processing owing to their intrinsically fast spin dynamics and vanishing net magnetization. Realizing this potential, however, requires understanding how terahertz fields excite coherent magnons and how the resulting spin motion is transduced into an optical signal. Here we report impulsive terahertz (THz) excitation and time-domain detection of the
antiferromagnetic resonance (AFMR) in single crystal manganese(II) oxide. Time-resolved birefringence measurements reveal long-lived coherent spin oscillations in the AFM phase. The resonance softens and becomes strongly damped upon warming toward the Néel temperature. 
Despite this similar
excitation behavior, the detected birefringence in MnO is markedly
weaker than in NiO. We associate this suppressed optical visibility
with the weak spin--orbit-mediated magneto-optical coupling of
orbital-singlet, high-spin Mn$^{2+}$. These results demonstrate that
coherent magnon excitation and its optical detection are governed by
distinct microscopic interactions.

%% Just tooling around with some different ideas / selling points -- Chris
%Antiferromagnets (AFMs) offer a promising platform for next-generation ultrafast information processing technologies, owing to their intrinsically fast spin dynamics and [???]. Realizing the potential of these THz scale magnetic excitations, however, requires an understanding of both the mechanisms by which terahertz fields excite coherent magnons and the time scales on which these modes decay. The canonical collinear AFM, manganese(II) oxide, provides an intriguing testbed to explore excitation mechanisms beyond linear electric-dipole and spin-orbit coupling.  Here we report impulsive terahertz (THz) excitation and time-domain detection of the antiferromagnetic resonance (AFMR) in single crystal MnO. Time-resolved birefringence measurements reveal long-lived coherent spin oscillations in the AFM phase when resonantly excited by the THz. These excitations soften and damp when approaching the N`{e}el temperature in a way inconsistent with a simple mean-field prediction but more consistent with [....].  We posit that the strongly suppressed THz excitation, relative to MnO's isostructural cousin nickel(II) oxide, is mediated by excitation through a magnetic-dipole Zeeman torque exerted by the THz magnetic field. These findings highlight the magnetic-dipole channel as an alternative mechanism for coherently driving spin dynamics in AFMs and establish time-domain access to the subsequent magnon decay.

\end{abstract}

% which is forbidden in MnO by inversion symmetry.  Unlike it's isostructural cousin, nickel(II) oxide, the $L=0$\ ground state of MnO lacks strong spin-orbit coupling 

\maketitle

% ============================================================
%\section{Introduction}
% ------------------------------------------------------------
Establishing efficient mechanisms for the generation, manipulation, and detection of collective spin excitations on sub-picosecond timescales is a central objective of ultrafast magnonics~\cite{kirilyuk2010,walowski2016}. Antiferromagnets (AFMs) are particularly attractive candidates for high frequency magnonic devices because their strong exchange interactions support spin dynamics in the terahertz (THz) frequency range, allowing high-bandwidth information transfer, and their lack of a net magnetic moment eliminates macroscopic stray fields~\cite{jungwirth2016,baltz2018,han2023coherent}. A central question, however, is how terahertz electromagnetic fields couple to the AFM order. Whereas the electric field can drive spin dynamics indirectly through spin--orbit-, orbital-, or lattice-mediated pathways~\cite{baierl2016,nova2017,mashkovich2021}, the magnetic field of a THz pulse can couple directly to the magnetic moments through the Zeeman interaction~\cite{kampfrath2011,baierl2016prl}. At sufficiently large amplitudes, coherent THz driving can access nonlinear spin motion and even ballistic switching~\cite{schlauderer2019}. 

\begin{figure*}[t]
\centering
\includegraphics[width=17.5cm]{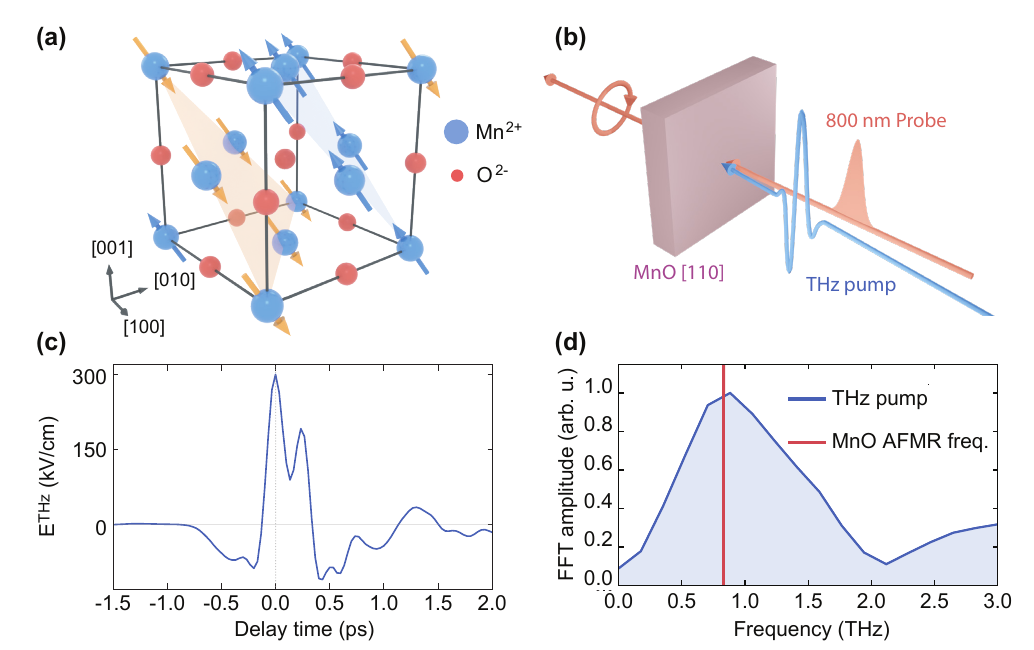}
\caption{\label{fig:overview} MnO magnetic structure and experimental configuration.
(a) Type-II antiferromagnetic order of the Mn$^{2+}$ sublattice in MnO. Mn$^{2+}$ spins are ferromagnetically aligned within $\{111\}$ planes, with adjacent planes stacked antiferromagnetically along the $[111]$ direction. The Mn$^{2+}$ spin orientations are indicated by arrows.
(b) Experimental schematic. High-field THz pulses generated in PNPA are incident on the $[110]$-oriented MnO crystal, and the transient birefringence response is detected through the resulting change in the polarization state of an $800$~nm probe. The THz electric field is independently characterized by electro-optic sampling in GaP.
(c) Temporal waveform of the PNPA-generated THz pump pulse.
(d) Corresponding Fourier spectrum of the THz pump. The low-temperature MnO AFMR frequency \(\omega_{\mathrm{AFMR}}=0.84\) THz is indicated by the vertical orange line.}
\end{figure*}

Manganese oxide (MnO) is a prototypical type-II AFM
insulator with a rock-salt structure. As shown in Fig.~\ref{fig:overview}(a), below the N\'eel
temperature $T_N \approx 118$~K, the Mn spins align ferromagnetically within $\{111\}$ planes, while neighboring planes stack antiferromagnetically along the $\langle111\rangle$ direction\cite{shull1949,roth1958,goodwin2006,keffer1957problem}. Together with NiO, MnO is one of the canonical transition-metal oxide antiferromagnets, with a well-established spin-wave dispersion~\cite{kittel2005,Pepy1974}. Its low-energy magnon dynamics, however, remain comparatively underexplored in the time domain.

As an easy-plane antiferromagnet, MnO supports two zone-center AFMR branches~\cite{keffer1952,mandel1973}. The high-frequency branch, corresponding predominantly to out-of-plane spin motion, occurs at $0.82$~THz at low temperature, whereas the low-frequency branch associated with the much weaker in-plane anisotropy lies in the microwave regime~\cite{sievers1963}. Importantly, the AFM transition in MnO preserves spatial inversion symmetry, forbidding direct linear electric-dipole excitation of the pure zone-center AFMR by parity. Instead, the zone-center magnon mode can couple directly to the THz magnetic field through the magnetic-dipole Zeeman interaction, while optical excitation may occur more weakly through the spin--orbit-mediated Raman tensor~\cite{fleury1968,chou1976}. 

As an important benchmark, terahertz-driven spin dynamics have been studied extensively in the isostructural AFM NiO, where intense THz magnetic-field transients coherently excite the $\sim 1$~THz magnon and produce nonlinear spin dynamics at higher fields~\cite{kampfrath2011,baierl2016prl}. Complementary optical studies have established coherent excitation of both AFMR branches and their polarization selection rules~\cite{satoh2010,higuchi2011,tzschaschel2017}. MnO provides a contrasting test case because the high-spin $3d^5$, $^6S$-like ground state of Mn$^{2+}$ is associated with weaker spin--orbit-mediated optical coupling and a correspondingly weaker coherent magneto-optical response. Previous work has observed the MnO AFMR as a resonant absorption in transmission THz time-domain spectroscopy~\cite{moriyasu2013}, while femtosecond optical excitation has been used to launch coherent magnons detected through THz emission~\cite{nishitani2013terahertz}. However, direct time-domain measurements of the coherent MnO response to strong THz excitation remain limited, leaving open how the excitation dynamics compare with the well-established behavior in NiO.

Here, we demonstrate, to our knowledge, the first phase-resolved strong-field THz excitation and optical time-domain readout of the MnO AFMR. This approach provides simultaneous access to the resonance frequency, damping, and phase of the coherent magnon relative to the driving THz field. The measured phase relative to the THz excitation closely resemble the response associated with Zeeman driving in NiO, supporting magnetic-dipole Zeeman coupling as the predominant excitation pathway under our experimental conditions. By contrast, the exceptionally weak birefringence response indicates inefficient optical transduction of the spin dynamics in MnO, consistent with its comparatively weak spin--orbit-mediated magneto-optical coupling.

% ============================================================
%\section{Experimental}
% ------------------------------------------------------------
Fig.~\ref{fig:overview}(b) shows a schematic of the THz pump--optical probe setup: high-field, single-cycle THz pulses are generated by pumping the organic nonlinear crystals PNPA~\cite{rovere2022} with $1300$~nm pulses from a three-stage optical parametric amplifier (OPA). Organic crystals provide efficient optical rectification and access to intense, broadband single-cycle THz fields~\cite{hauri2011,jazbinsek2019}. The OPA is driven by a multipass Ti:sapphire amplifier operating at $350$~Hz with a pulse duration of $\sim50$~fs and delivers pulse energies of up to $2$~mJ. The resulting THz waveform is characterized by standard free-space electro-optic sampling (EOS)~\cite{wu1995} in a $100$~\textmu m-thick GaP crystal using balanced detection, with a peak electric field of approximately $300$~kV/cm; the measured waveform and spectrum are shown in Figs.~\ref{fig:overview}(c) and \ref{fig:overview}(d). The broadband THz pump spectrum spans the MnO AFMR frequency.

\begin{figure*}[t]
\centering
\includegraphics[width=\linewidth]{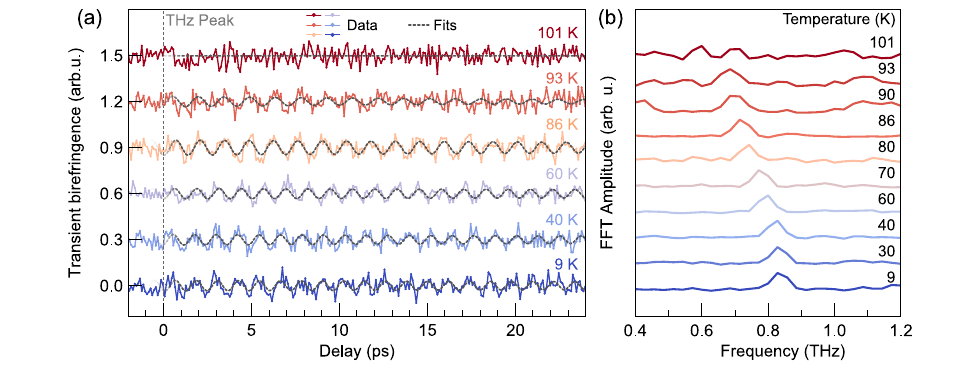}
\caption{\label{fig:timedomain} Temperature dependence of the MnO AFMR.
(a) Time-domain transient birefringence of MnO at selected temperatures (colored lines), vertically offset for clarity. Single-mode damped-sinusoidal fits are overlaid as black dashed lines.  No resolved oscillation is observed at $101$~K. The vertical dashed line marks the peak of the incident THz field, defined as $t=0$ for all traces.
(b) Fourier spectra of the time-domain transients at selected temperatures, vertically offset for clarity. The AFMR peak shifts to lower frequency with increasing temperature. No distinct AFMR peak is resolved at $101$~K.}
\end{figure*}
The sample is a $5\times5$~mm$^2$, $100$~\textmu m-thick [110]-oriented MnO single crystal mounted in a closed-cycle helium cryostat. Measurements are performed in transmission geometry between $7$ and $120$~K. The coherent spin dynamics are detected through the pump-induced change in optical birefringence using an $800$~nm probe in a balanced detection scheme, with the THz pump and optical probe co-polarized and normally incident on the sample. The probe delay is scanned in $0.1$~ps steps over a $40$~ps time window, and the signal is detected by a lock-in amplifier. For phase-sensitive measurements, all scans are referenced to a common delay-stage origin, preserving the relative timing between the incident THz field and the coherent magnon response across all temperatures studied.

% ============================================================
%\section{Results}
% ------------------------------------------------------------

\begin{figure}[t]
\centering
\includegraphics[width=\linewidth]{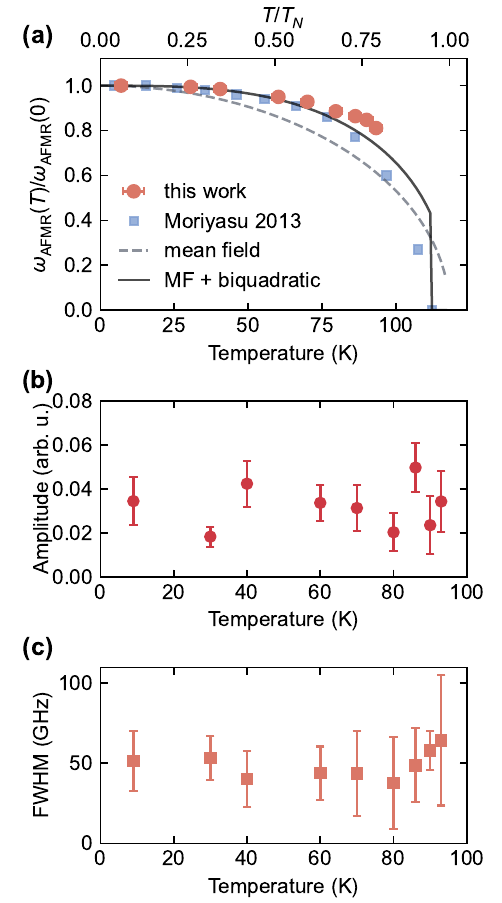}
\caption{\label{fig:freqfft} Temperature dependence of the antiferromagnetic resonance in MnO. (a) Normalized AFMR frequency, \(\omega_{\mathrm{AFMR}}(T)/\omega_{\mathrm{AFMR}}(0)\), plotted as a function of temperature (bottom axis) and normalized temperature \(T/T_N\) (top axis), with \(T_N=118\) K and \(\omega_{\mathrm{AFMR}}(0)=0.84\) THz. Red circles show the present measurements, while blue squares show data from Moriyasu et al. (2013)~\cite{moriyasu2013}. The dashed gray and solid black curves represent the mean-field and mean-field-plus-biquadratic models reported in the same reference, respectively. (b) Fitted amplitude of the coherent AFMR oscillation as a function of temperature. (c) AFMR linewidth, expressed as the full width at half maximum (FWHM), obtained from fits to the Fourier spectra. Error bars indicate 95\% confidence intervals.}
\end{figure}

Fig.~\ref{fig:timedomain}(a) shows the time-domain response of MnO following THz excitation at selected temperatures. Well below $T_N$, the signal exhibits a clear, long-lived coherent oscillation that is well described by a single exponentially damped sinusoid,
$S(t)=A e^{-t/\tau}\cos(2\pi f t+\varphi)$.
The oscillation frequency decreases with increasing temperature, while the fitted amplitude remains approximately constant over most of the measured temperature range (Fig.\ref{fig:freqfft}(b)). Fig.~\ref{fig:timedomain}(b) shows the corresponding Fourier spectra. Below $\sim80$~K, the AFMR appears as a narrow resonance near $0.8$~THz. Upon further warming, the resonance shifts to lower frequency and broadens, and is no longer resolved at $101$~K.

The extracted resonance frequency is summarized in Fig.~\ref{fig:freqfft}(a). At base temperature, the AFMR occurs at $f=0.83$~THz, in close agreement with previous far-infrared and THz-TDS measurements~\cite{sievers1963,moriyasu2013}. The frequency remains nearly constant up to $\sim60$~K and then progressively decreases, reaching $0.68$~THz at $93$~K. This temperature dependence deviates from the simple mean-field form $\sqrt{1-(T/T_N)^2}$ and is instead consistent with the molecular-field model incorporating biquadratic exchange~\cite{moriyasu2013}.

The temperature dependence is also pronounced in the magnon damping. Below $70$~K, the coherent oscillation persists beyond the measured time window, yielding a lower bound of $\tau\gtrsim80$~ps for the lifetime. The decay becomes resolvable near $86$~K, where $\tau=76$~ps, and increases rapidly upon further warming, with the lifetime decreasing to $\tau=20$~ps at $93$~K. The corresponding linewidths are summarized in Fig.~\ref{fig:freqfft}(c). Measurements with finer temperature spacing between $93$ and $101$~K would be required to resolve the disappearance of the coherent response in greater detail.

\begin{figure}[t]
\centering
\includegraphics[width=\linewidth]{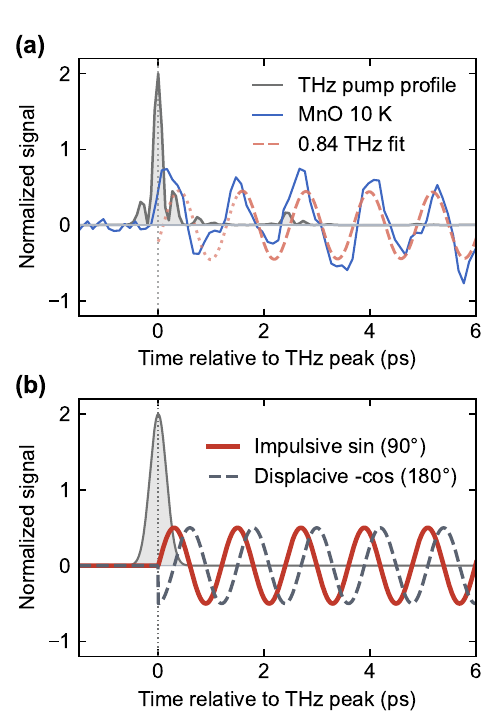}
\caption{\label{fig:phase} Excitation phase of the MnO AFMR at $10$~K under THz excitation.
(a) Time-domain magnon response (blue) of MnO and fit to a damped sinusoid (orange dashed line), shown together with the THz pump envelope (gray). Time is referenced to the peak of the THz field ($t=0$). The fitted phase is $\varphi \approx 29^\circ$.
(b) Comparison of the magnon oscillations corresponding to an impulsive sine-like response ($90^\circ$, solid red) and a displacive cosine-like response ($180^\circ$, dashed gray).}
\end{figure}

% ============================================================
%\section{Discussion}
%
The observed mode is identified as the high-frequency AFMR branch of MnO from its resonance frequency and temperature dependence. The base-temperature frequency agrees closely with previous far-infrared~\cite{sievers1963} and THz-TDS~\cite{moriyasu2013} measurements, while its temperature-dependent softening is well described by a molecular-field model incorporating biquadratic exchange. This behavior closely parallels that of the corresponding AFMR mode in NiO~\cite{kampfrath2011}, although the detected birefringence response in MnO is substantially weaker. Importantly, the small optical signal does not necessarily imply a small spin-precession amplitude. Instead, the weak magneto-optical response is consistent with the electronic structure of Mn$^{2+}$: its high-spin $3d^5$, $^6S$ ground state is an orbital singlet ($L=0$), for which single-ion anisotropy and spin--orbit-mediated optical coupling are expected to be weak~\cite{terakura1984,schroen2012,uchiyama2012}.

The symmetry of MnO strongly constrains the possible coupling channels between the THz field and the AFMR. The rock-salt lattice is centrosymmetric, and the type-II antiferromagnetic order preserves spatial inversion because the Mn spins transform as axial vectors. Direct linear electric-dipole excitation of the zone-center single-magnon mode is therefore forbidden by parity. By contrast, the THz magnetic field can couple directly to the AFMR through the magnetic-dipole Zeeman interaction, as established for NiO~\cite{kampfrath2011,baierl2016prl}. One-magnon Raman excitation is also symmetry allowed through the antisymmetric, spin--orbit-mediated Raman tensor~\cite{fleury1968}, but is expected to be weak for orbital-singlet Mn$^{2+}$~\cite{chou1976,grimsditch1998}. Additional electric-field-driven pathways, including higher-order magnon processes, transient anisotropy, and spin--lattice coupling, may also contribute indirectly~\cite{baierl2016,nova2017,mashkovich2021}; however, these mechanisms do not constitute a direct linear electric-dipole coupling to the zone-center AFMR and cannot be quantitatively excluded by the present measurements.

The excitation phase provides an additional constraint on the driving mechanism. Referencing all scans to a common delay-stage origin allows the coherent magnon phase to be compared directly with the THz field. At $10$~K, the fitted oscillation yields $\varphi \approx 29^\circ$ (Fig.~\ref{fig:phase}(a)), placing the response closer to the sine-like impulsive limit than to the cosine-like displacive limit shown in Fig.~\ref{fig:phase}(b). Such a response is consistent with a Zeeman torque,
$\bm{\tau}=\gamma\,\mathbf{M}\times\mathbf{H}_{\mathrm{THz}}$,
acting during the sub-picosecond THz magnetic-field transient and imparting an initial momentum to the magnon coordinate. In contrast, a displacive mechanism, such as field rectification or a sudden change in magnetic anisotropy, would produce a more cosine-like response associated with a transient shift of the equilibrium configuration~\cite{garrett1996coherent}.

To establish the microscopic origin of the observed antiferromagnetic resonance in MnO, we used a first-principles spin model including magnetic exchange interactions, magnetocrystalline anisotropy, and long-range magnetic dipole--dipole interactions. The \textit{ab initio} and model details can be found in Refs. \cite{tong2026direct, He2021TB2J} By integrating \textit{ab initio} calculations with the Heisenberg spin Hamiltonian~\cite{lines1965antiferromagnetism,korotin2015calculation,dossantos2026}, we calculated the microscopic magnetic exchange interactions for the high-spin $S = 5/2$ Mn, yielding first-nearest-neighbor couplings of $J_1^+ = -0.971$~meV and $J_1^- = -0.647$~meV, alongside a next-nearest-neighbor coupling of $J_2 = -0.892$~meV. The $J_{1}^{+(-)}$ parameters govern the nearest-neighbor Mn--Mn magnetic coupling, differentiating between Mn ion pairs within the same ferromagnetic sheet that share parallel spin orientations ($J_{1}^{+}$) and pairs across adjacent sheets with antiparallel alignments ($J_{1}^{-}$). These $J_{1}$ pathways emerge from a delicate competition between direct $3d$--$3d$ orbital overlap and $90^{\circ}$ Mn--O--Mn super-exchange mechanisms, where electrons interact via intermediate O atoms. In contrast, the robust next-nearest-neighbor interaction $J_{2}$ is mediated by a linear $180^{\circ}$ Mn--O--Mn bond configuration~\cite{Binci2025,dossantos2026}, functioning as a strong antiferromagnetic super-exchange channel that stabilizes the magnetic ground state and dictates the broader spin ordering. Noncollinear spin--orbit coupling calculations identify the $(111)$ plane as the easy magnetization plane, yielding a magnetocrystalline anisotropy energy of $19.6~\mu$eV/Mn. In contrast, a converged lattice sum of the full magnetic dipole--dipole tensor gives an interaction energy of $0.267$~meV/Mn, demonstrating that dipolar interactions dominate the restoring torque for out-of-plane motion, consistent with the known role of dipolar interactions in lifting the zone-center magnon degeneracy~\cite{Pepy1974}. Linearizing $\mathcal{H} = \mathcal{H}_{\text{ex}} + \mathcal{H}_{\text{SOC}} + \mathcal{H}_{\text{dip}}$ about the type-II antiferromagnetic ground state yields two zone-center modes at $0.122$ and $0.866$~THz. The upper mode is $99.6\%$ polarized along $[111]$, and its calculated energy of $3.58$~meV agrees within $\sim 4\%$ with the observed $3.45$~meV ($0.833$~THz) resonance. Because its eigenvector contains a small, in-phase transverse component of the two Mn sublattices, it produces a finite oscillating net magnetization that provides a magnetic-dipole coupling channel to the THz magnetic field. These results reveal that the gap of the observed resonance is governed primarily by long-range dipolar interactions rather than spin–orbit coupling alone.

% ============================================================
%\section{Conclusion}
%
In summary, we have demonstrated impulsive terahertz-pump excitation and time-domain readout
of the antiferromagnetic resonance in single-crystal MnO, resolving its frequency, lifetime, and excitation phase within the AFM order. The AFMR softens from $0.83$~THz at low temperature to $0.68$~THz at $93$~K, with a temperature dependence that deviates from simple mean-field behavior and is well described by a molecular-field model including biquadratic exchange. The coherent response becomes strongly damped upon warming and is no longer resolved near $T/T_N\approx0.86$. Its sine-like excitation phase is consistent with a predominantly linear Zeeman torque from the THz magnetic field. The exceptionally weak birefringence response is instead attributed to inefficient spin--orbit-mediated magneto-optical transduction in orbital-singlet, high-spin Mn$^{2+}$. These results show that coherent antiferromagnetic excitation and its optical visibility can be governed by distinct microscopic interactions.

\begin{acknowledgments}
This work was supported as part of the Center for Energy Efficient Magnonics, an Energy Frontier Research Center funded by the U.S. Department of Energy, Office of Science, Basic Energy Sciences at SLAC National Accelerator Laboratory under Contract No. DE-AC02-76SF00515. D.P. and H.P. acknowledge the use of the computational facilities on the Frontera supercomputer at the Texas Advanced Computing Center (TACC) through Pathways allocation DMR23051.
\end{acknowledgments}

\section*{Author Declarations}
\noindent\textbf{Conflict of Interest.} The authors have no conflicts to disclose.

\section*{Data Availability}
The data that support the findings of this study are available from the
corresponding author upon reasonable request.

\bibliography{references}

@article{tong2026direct,
  author    = {Tong, Junwei and Paudyal, Hari and Liu, Xiangcheng and Takana, Lerato and Mikhailova, Katya and Suzuki, Yuri and Paudyal, Durga and Li, Xiaoqin},
  title     = {Direct Observation of Tunable Magnons in Epitaxial Lithium Aluminum Ferrite Thin Films},
  journal   = {Nano Letters},
  volume    = {26},
  number    = {9},
  pages     = {3073--3079},
  year      = {2026},
  doi       = {10.1021/acs.nanolett.5c05922},
  url       = {https://pubs.acs.org/doi/abs/10.1021/acs.nanolett.5c05922}
}

@article{He2021TB2J,
  title     = {TB2J: A python package for computing magnetic interaction parameters},
  author    = {Xu He and Nicole Helbig and Matthieu J. Verstraete and Eric Bousquet},
  journal   = {Computer Physics Communications},
  volume    = {264},
  pages     = {107938},
  year      = {2021},
  doi       = {10.1016/j.cpc.2021.107938},
  url       = {https://www.sciencedirect.com/science/article/pii/S0010465521000679},
  eprint    = {2009.01910},
  archivePrefix = {arXiv},
  primaryClass  = {cond-mat.mtrl-sci}
}

@article{kirilyuk2010,
  author  = {Kirilyuk, Andrei and Kimel, Alexey V. and Rasing, Theo},
  title   = {Ultrafast optical manipulation of magnetic order},
  journal = {Rev. Mod. Phys.},
  volume  = {82},
  pages   = {2731--2784},
  year    = {2010},
  doi     = {10.1103/RevModPhys.82.2731}
}

@article{jungwirth2016,
  author  = {Jungwirth, T. and Marti, X. and Wadley, P. and Wunderlich, J.},
  title   = {Antiferromagnetic spintronics},
  journal = {Nat. Nanotechnol.},
  volume  = {11},
  pages   = {231--241},
  year    = {2016},
  doi     = {10.1038/nnano.2016.18}
}

@article{baltz2018,
  author  = {Baltz, V. and Manchon, A. and Tsoi, M. and Moriyama, T. and Ono, T. and Tserkovnyak, Y.},
  title   = {Antiferromagnetic spintronics},
  journal = {Rev. Mod. Phys.},
  volume  = {90},
  pages   = {015005},
  year    = {2018},
  doi     = {10.1103/RevModPhys.90.015005}
}

@article{walowski2016,
  author  = {Walowski, Jakob and M{\"u}nzenberg, Markus},
  title   = {Perspective: Ultrafast magnetism and {THz} spintronics},
  journal = {J. Appl. Phys.},
  volume  = {120},
  pages   = {140901},
  year    = {2016},
  doi     = {10.1063/1.4958846}
}

@article{keffer1952,
  author  = {Keffer, F. and Kittel, C.},
  title   = {Theory of Antiferromagnetic Resonance},
  journal = {Phys. Rev.},
  volume  = {85},
  pages   = {329--337},
  year    = {1952},
  doi     = {10.1103/PhysRev.85.329}
}

@article{shull1949,
  author  = {Shull, C. G. and Smart, J. Samuel},
  title   = {Detection of Antiferromagnetism by Neutron Diffraction},
  journal = {Phys. Rev.},
  volume  = {76},
  pages   = {1256--1257},
  year    = {1949},
  doi     = {10.1103/PhysRev.76.1256}
}

@article{roth1958,
  author  = {Roth, W. L.},
  title   = {Magnetic Structures of {MnO}, {FeO}, {CoO}, and {NiO}},
  journal = {Phys. Rev.},
  volume  = {110},
  pages   = {1333--1341},
  year    = {1958},
  doi     = {10.1103/PhysRev.110.1333}
}

@article{goodwin2006,
  author  = {Goodwin, Andrew L. and Tucker, Matthew G. and Dove, Martin T. and Keen, David A.},
  title   = {Magnetic Structure of {MnO} at 10 K from Total Neutron Scattering Data},
  journal = {Phys. Rev. Lett.},
  volume  = {96},
  pages   = {047209},
  year    = {2006},
  doi     = {10.1103/PhysRevLett.96.047209}
}

@article{sievers1963,
  author  = {Sievers, A. J. and Tinkham, M.},
  title   = {Far Infrared Antiferromagnetic Resonance in {MnO} and {NiO}},
  journal = {Phys. Rev.},
  volume  = {129},
  pages   = {1566--1571},
  year    = {1963},
  doi     = {10.1103/PhysRev.129.1566}
}

@article{mandel1973,
  author  = {Mandel', V. S. and Voronkov, V. D. and Gromzin, D. E.},
  title   = {Antiferromagnetic resonance in {MnO}},
  journal = {Sov. Phys. JETP},
  volume  = {36},
  pages   = {521--524},
  year    = {1973}
}

@article{Pepy1974,
  author  = {Pepy, G.},
  title   = {Spin waves in {MnO}: From 4 K to temperatures close to {$T_N$}},
  journal = {J. Phys. Chem. Solids},
  volume  = {35},
  pages   = {433--444},
  year    = {1974},
  doi     = {10.1016/S0022-3697(74)80037-5}
}

@book{kittel2005,
  author    = {Kittel, Charles},
  title     = {Introduction to Solid State Physics},
  edition   = {8th},
  publisher = {Wiley},
  address   = {Hoboken, NJ},
  year      = {2005},
  pages     = {341}
}

@article{fleury1968,
  author  = {Fleury, P. A. and Loudon, R.},
  title   = {Scattering of Light by One- and Two-Magnon Excitations},
  journal = {Phys. Rev.},
  volume  = {166},
  pages   = {514--530},
  year    = {1968},
  doi     = {10.1103/PhysRev.166.514}
}

@article{chou1976,
  author  = {Chou, H.-H. and Fan, H. Y.},
  title   = {Light scattering by magnons in {CoO}, {MnO}, and {$\alpha$-MnS}},
  journal = {Phys. Rev. B},
  volume  = {13},
  pages   = {3924--3938},
  year    = {1976},
  doi     = {10.1103/PhysRevB.13.3924}
}

@article{kampfrath2011,
  author  = {Kampfrath, T. and Sell, A. and Klatt, G. and Pashkin, A. and M{\"a}hrlein, S. and Dekorsy, T. and Wolf, M. and Fiebig, M. and Leitenstorfer, A. and Huber, R.},
  title   = {Coherent terahertz control of antiferromagnetic spin waves},
  journal = {Nat. Photonics},
  volume  = {5},
  pages   = {31--34},
  year    = {2011},
  doi     = {10.1038/nphoton.2010.259}
}

@article{satoh2010,
  author  = {Satoh, Takuya and Cho, Sung-Jin and Iida, Ryugo and Shimura, Tsutomu and Kuroda, Kazuo and Ueda, Hiroaki and Ueda, Yutaka and Ivanov, B. A. and Nori, Franco and Fiebig, Manfred},
  title   = {Spin Oscillations in Antiferromagnetic {NiO} Triggered by Circularly Polarized Light},
  journal = {Phys. Rev. Lett.},
  volume  = {105},
  pages   = {077402},
  year    = {2010},
  doi     = {10.1103/PhysRevLett.105.077402}
}

@article{higuchi2011,
  author  = {Higuchi, Takuya and Kanda, Natsuki and Tamaru, Hiroharu and Kuwata-Gonokami, Makoto},
  title   = {Selection Rules for Light-Induced Magnetization of a Crystal with Threefold Symmetry: The Case of Antiferromagnetic {NiO}},
  journal = {Phys. Rev. Lett.},
  volume  = {106},
  pages   = {047401},
  year    = {2011},
  doi     = {10.1103/PhysRevLett.106.047401}
}

@article{tzschaschel2017,
  author  = {Tzschaschel, Christian and Otani, Kensuke and Iida, Ryugo and Shimura, Tsutomu and Ueda, Hiroaki and G{\"u}nther, Stefan and Fiebig, Manfred and Satoh, Takuya},
  title   = {Ultrafast optical excitation of coherent magnons in antiferromagnetic {NiO}},
  journal = {Phys. Rev. B},
  volume  = {95},
  pages   = {174407},
  year    = {2017},
  doi     = {10.1103/PhysRevB.95.174407}
}

@article{baierl2016prl,
  author  = {Baierl, S. and Mentink, J. H. and Hohenleutner, M. and Braun, L. and Do, T.-M. and Lange, C. and Sell, A. and Fiebig, M. and Woltersdorf, G. and Kampfrath, T. and Huber, R.},
  title   = {Terahertz-Driven Nonlinear Spin Response of Antiferromagnetic Nickel Oxide},
  journal = {Phys. Rev. Lett.},
  volume  = {117},
  pages   = {197201},
  year    = {2016},
  doi     = {10.1103/PhysRevLett.117.197201}
}

@article{baierl2016,
  author  = {Baierl, S. and Hohenleutner, M. and Kampfrath, T. and Zvezdin, A. K. and Kimel, A. V. and Huber, R. and Mikhaylovskiy, R. V.},
  title   = {Nonlinear spin control by terahertz-driven anisotropy fields},
  journal = {Nat. Photonics},
  volume  = {10},
  pages   = {715--718},
  year    = {2016},
  doi     = {10.1038/nphoton.2016.181}
}

@article{nova2017,
  author  = {Nova, T. F. and Cartella, A. and Cantaluppi, A. and F{\"o}rst, M. and Bossini, D. and Mikhaylovskiy, R. V. and Kimel, A. V. and Merlin, R. and Cavalleri, A.},
  title   = {An effective magnetic field from optically driven phonons},
  journal = {Nat. Phys.},
  volume  = {13},
  pages   = {132--136},
  year    = {2017},
  doi     = {10.1038/nphys3925}
}

@article{mashkovich2021,
  author  = {Mashkovich, E. A. and Grishunin, K. A. and Dubrovin, R. M. and others},
  title   = {Terahertz light--driven coupling of antiferromagnetic spins to lattice},
  journal = {Science},
  volume  = {374},
  pages   = {1608--1611},
  year    = {2021},
  doi     = {10.1126/science.abk1121}
}

@article{schlauderer2019,
  author  = {Schlauderer, S. and Lange, C. and Baierl, S. and others},
  title   = {Temporal and spectral fingerprints of ultrafast all-coherent spin switching},
  journal = {Nature},
  volume  = {569},
  pages   = {383--387},
  year    = {2019},
  doi     = {10.1038/s41586-019-1174-7}
}

@article{moriyasu2013,
  author  = {Moriyasu, T. and Wakabayashi, S. and Kohmoto, T.},
  title   = {Observation of Antiferromagnetic Magnons and Magnetostriction in Manganese Oxide Using Terahertz Time-Domain Spectroscopy},
  journal = {J. Infrared Millim. Terahertz Waves},
  volume  = {34},
  pages   = {277--283},
  year    = {2013},
  doi     = {10.1007/s10762-013-9963-9}
}

@article{nishitani2013terahertz,
  title={Terahertz radiation from antiferromagnetic MnO excited by optical laser pulses},
  author={Nishitani, Junichi and Nagashima, Takeshi and Hangyo, Masanori},
  journal={Applied Physics Letters},
  volume={103},
  number={8},
  year={2013},
  publisher={AIP Publishing}
}

@article{rovere2022,
  author  = {Rovere, A. and Jeong, Y.-G. and Piccoli, R. and others},
  title   = {A new standard in high-field terahertz generation: The organic nonlinear optical crystal {PNPA}},
  journal = {ACS Photonics},
  volume  = {9},
  pages   = {3720--3726},
  year    = {2022},
  doi     = {10.1021/acsphotonics.2c01336}
}

@article{hauri2011,
  author  = {Hauri, Christoph P. and Ruchert, Clemens and Vicario, Carlo and Ardana, Fernando},
  title   = {Strong-field single-cycle {THz} pulses generated in an organic crystal},
  journal = {Appl. Phys. Lett.},
  volume  = {99},
  pages   = {161116},
  year    = {2011},
  doi     = {10.1063/1.3655331}
}

@article{jazbinsek2019,
  author  = {Jazbinsek, Mojca and Puc, Uros and Abina, Andreja and Zidansek, Aleksander},
  title   = {Organic Crystals for {THz} Photonics},
  journal = {Appl. Sci.},
  volume  = {9},
  pages   = {882},
  year    = {2019},
  doi     = {10.3390/app9050882}
}

@article{wu1995,
  author  = {Wu, Q. and Zhang, X.-C.},
  title   = {Free-space electro-optic sampling of terahertz beams},
  journal = {Appl. Phys. Lett.},
  volume  = {67},
  pages   = {3523--3525},
  year    = {1995},
  doi     = {10.1063/1.114909}
}

@article{grimsditch1998,
  author  = {Grimsditch, M. and McNeil, L. E. and Lockwood, D. J.},
  title   = {Unexpected behavior of the antiferromagnetic mode of {NiO}},
  journal = {Phys. Rev. B},
  volume  = {58},
  pages   = {14462--14466},
  year    = {1998},
  doi     = {10.1103/PhysRevB.58.14462}
}

@article{terakura1984,
  author  = {Terakura, K. and Oguchi, T. and Williams, A. R. and K{\"u}bler, J.},
  title   = {Band theory of insulating transition-metal monoxides: Band-structure calculations},
  journal = {Phys. Rev. B},
  volume  = {30},
  pages   = {4734--4747},
  year    = {1984},
  doi     = {10.1103/PhysRevB.30.4734}
}

@article{schroen2012,
  author  = {Schr{\"o}n, A. and R{\"o}dl, C. and Bechstedt, F.},
  title   = {Crystalline and magnetic anisotropy of the 3$d$-transition metal monoxides {MnO}, {FeO}, {CoO}, and {NiO}},
  journal = {Phys. Rev. B},
  volume  = {86},
  pages   = {115134},
  year    = {2012},
  doi     = {10.1103/PhysRevB.86.115134}
}

@article{uchiyama2012,
  author  = {Uchiyama, H.},
  title   = {Potential asymmetry in antiferromagnetic 3$d$ transition metal monoxides},
  journal = {Phys. Rev. B},
  volume  = {85},
  pages   = {014419},
  year    = {2012},
  doi     = {10.1103/PhysRevB.85.014419}
}

@article{korotin2015calculation,
  author  = {Korotin, D. M. and Mazurenko, V. V. and Anisimov, V. I. and Streltsov, S. V.},
  title   = {Calculation of exchange constants of the Heisenberg model in plane-wave-based methods using the Green's function approach},
  journal = {Phys. Rev. B},
  volume  = {91},
  pages   = {224405},
  year    = {2015},
  doi     = {10.1103/PhysRevB.91.224405}
}

@article{Binci2025,
  author  = {Binci, Leonardo and Marzari, Nicola and Timrov, Iurii},
  title   = {Magnons from time-dependent density-functional perturbation theory and nonempirical Hubbard functionals},
  journal = {npj Comput. Mater.},
  volume  = {11},
  pages   = {100},
  year    = {2025},
  doi     = {10.1038/s41524-025-01570-0}
}

@article{dossantos2026,
  author  = {dos Santos, Flaviano Jos{\'e} and Binci, Luca and Menichetti, Guido and Mahajan, Ruchika and Marzari, Nicola and Timrov, Iurii},
  title   = {Comparative study of magnetic exchange parameters and magnon dispersions in {NiO} and {MnO} from first principles},
  journal = {Phys. Rev. B},
  volume  = {113},
  pages   = {024427},
  year    = {2026},
  doi     = {10.1103/gtxm-6vtg}
}

@article{lines1965antiferromagnetism,
  title={Antiferromagnetism in the face-centered cubic lattice. II. Magnetic properties of MnO},
  author={Lines, ME and Jones, ED},
  journal={Physical Review},
  volume={139},
  number={4A},
  pages={A1313},
  year={1965},
  publisher={APS}
}

@article{han2023coherent,
  title={Coherent antiferromagnetic spintronics},
  author={Han, Jiahao and Cheng, Ran and Liu, Luqiao and Ohno, Hideo and Fukami, Shunsuke},
  journal={Nature Materials},
  volume={22},
  number={6},
  pages={684--695},
  year={2023},
  publisher={Nature Publishing Group UK London}
}

@article{keffer1957problem,
  title={Problem of spin arrangements in MnO and similar antiferromagnets},
  author={Keffer, F and O'sullivan, W},
  journal={Physical Review},
  volume={108},
  number={3},
  pages={637},
  year={1957},
  publisher={APS}
}

@article{garrett1996coherent,
  title={Coherent THz phonons driven by light pulses and the Sb problem: What is the mechanism?},
  author={Garrett, Gregory A and Albrecht, TF and Whitaker, JF and Merlin, Roberto},
  journal={Physical review letters},
  volume={77},
  number={17},
  pages={3661},
  year={1996},
  publisher={APS}
}

\end{document}